\documentclass[twocolumn,aps,floatfix,showpacs,bibnotes,prl]{revtex4-1}
\usepackage{graphicx}% Include figure files
\usepackage{amsmath}
\usepackage{bm}% bold math
\begin{document}

\title{Measurement of the mobility edge for 3D Anderson localization}
\author{G. Semeghini*$^1$, M. Landini*$^{1,2}$, P. Castilho$^{1 \dag}$, S. Roy$^1$, G. Spagnolli$^{1,2}$, A. Trenkwalder$^{1,2}$, M. Fattori$^{1,2,3}$, M. Inguscio$^{1,4}$ \& G. Modugno$^{1,2,3}$}

\affiliation{ $^{1}$LENS and Dipartimento di Fisica e Astronomia, Universit\'a di Firenze, Via N. Carrara 1, 50019, Sesto Fiorentino, Italy}
\affiliation{ $^{2}$CNR-INO, Via G. Sansone 1, 50019, Sesto Fiorentino, Italy}
\affiliation{ $^{3}$INFN - Sezione di Firenze, Via G. Sansone 1, 50019, Sesto Fiorentino, Italy}
\affiliation{ $^{4}$INRIM, Strada delle Cacce 91, 10135, Torino, Italy}
\affiliation{ $^{\dag}$\textnormal{Present address}: Instituto de F\'isica de S\~ao Carlos, Universidade de S\~ao Paulo, C.P. 369, 13560-970, S\~ao Carlos, SP, Brazil}

\maketitle

\textbf{Anderson localization is a universal phenomenon affecting non-interacting quantum particles in disorder. In three spatial dimensions it becomes particularly interesting to study because of the presence of a quantum phase transition from localized to extended states, predicted by P.W. Anderson in his seminal work \cite{Anderson}, taking place at a critical energy, the so-called mobility edge. The possible relation of the Anderson transition to the metal-insulator transitions observed in materials \cite{Lee} has originated a flurry of theoretical studies during the past 50 years \cite{Lee,Thouless,Abrahams,Vollhardt,Kramer,Mirlin,Review}, and it is now possible to predict very accurately the mobility edge starting from models of the microscopic disorder \cite{Slevin}. However, the experiments performed so far with photons, ultrasound and ultracold atoms, while giving evidence of the transition \cite{Hu, Sperling, Chabe, Kondov, Jendrzejewski}, could not provide a precise measurement of the mobility edge. In this work we are able to obtain such a measurement using an ultracold atomic system in a disordered speckle potential, thanks to a precise control of the system energy. We find that the mobility edge is close to the mean disorder energy at small disorder strengths, while a clear effect of the spatial correlation of the disorder appears at larger strengths. The precise knowledge of the disorder properties in our system offers now the opportunity for an unprecedented experiment-theory comparison for 3D Anderson localization, which is also a necessary step to start the exploration of novel regimes for many-body disordered systems.}

The initial theoretical work by Anderson \cite{Anderson} was motivated by the observation of localization phenomena in solid-state systems \cite{Feher}. Studying a specific model of a disordered lattice, he realized that a critical amount of disorder could lead to a localization transition for electrons, due to quantum interference on the wavefunction. The possible connection between the Anderson transition and metal-insulator transitions \cite{Mott} started a huge theoretical investigation, which eventually led to a consensus between numerical \cite{Thouless, Kramer, Mirlin, Slevin, Review} and analytical results \cite{Abrahams, Vollhardt, Lee, Review} about the mobility edge and the critical properties at the transition. However, these results cannot be tested on electronic systems, where interactions turn the single-particle Anderson problem into a much more complex one \cite{Review}. The universal nature of Anderson localization allows to study some aspects of these phenomena with sound and light waves \cite{Hu, Sperling}, which have provided a test of the Ioffe-Regel criterion \cite{Ioffe}, or with atomic kicked rotors \cite{Chabe}, which have been employed to measure the critical properties of the transition. Ultracold atoms in disordered optical potentials can realize a system with full control on the microscopic Hamiltonian as in the Anderson theory \cite{Lewenstein, Billy, Roati1D}. However, recent experiments \cite{Kondov, Jendrzejewski}, while demonstrating the occurrence of an Anderson transition, could not precisely locate the mobility edge because of the difficulty in controlling the system energy.
We now develop a novel method to control the energy of an ultracold atomic system in disorder, and we use it to measure precisely the mobility edge and its dependence on the disorder strength. To realize the disorder we employ an optical speckle potential, for which the intensity distribution and spatial correlations can be precisely measured. Our energy-control strategy is based on three key parts. First, we achieve fully-localized, low-energy samples by loading a Bose-Einstein condensate almost adiabatically into the disorder, thanks to a slow cancellation of the atom-atom interactions. Then, we estimate the actual energy distribution of the samples by combining measurements of their kinetic energy with numerical simulations of the low energy eigenstates. Finally, we use a time-dependent modulation of the disorder to produce controlled excitations and we deduce the mobility edge from the measurement of the energy needed to break localization.

\begin{figure}
\includegraphics{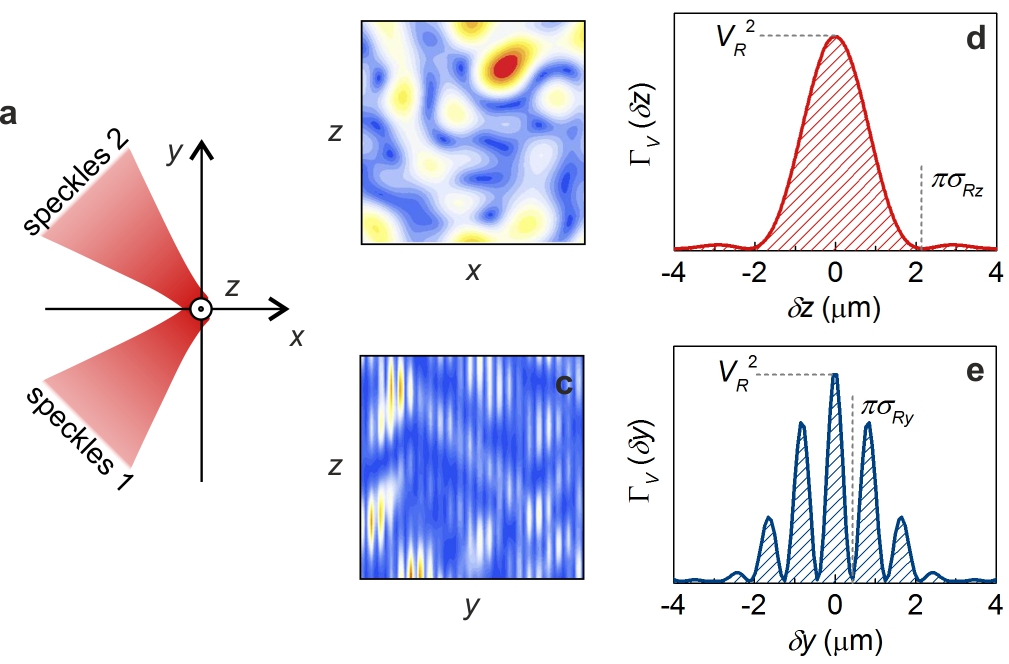}
\label{fig1}
\caption{\textbf{3D speckle disorder}. \textbf{a}) Sketch of the speckles geometry. \textbf{b,c}) Disordered potential calculated along two principal planes (12.5$\mu m$x12.5$\mu m$). \textbf{d,e}) Calculated intensity autocorrelation functions along two principal axes. Along $z$ and $x$, $\Gamma_V$ is a Bessel-type function, while along $y$ an extra modulation comes from interference.}
\end{figure}

The ultracold sample is composed by about 10$^5$ $^{39}$K atoms, for which the interaction can be controlled via a Feshbach resonance \cite{Roati}. The sample is initially cooled down to Bose-Einstein condensation with repulsive interaction in a harmonic trap. The disorder is generated by two coherent speckle fields \cite{Jendrzejewski} that are blue-detuned from the atomic transitions, hence producing a repulsive potential. As sketched in Fig.1, the two speckles cross each other at 90$^{\circ}$ with parallel linear polarizations, creating a 3D intensity distribution with short correlation lengths along all directions, whose geometric average is $\pi\sigma_R$=1.3$\mu$m. The two relevant energy scales are the disorder strength $V_R$, which represents both the mean value of the potential and its standard deviation, and the correlation energy $E_R=\hbar^2/m \sigma_R^2=$73nK. $V_R$ can be controlled via the total speckles power, and it is accurately calibrated. The two speckles envelopes are Gaussian with waists of about 1300$\mu$m, hence much wider than the typical atomic distributions, ensuring the homogeneity of $V_R$.

\begin{figure}
\includegraphics{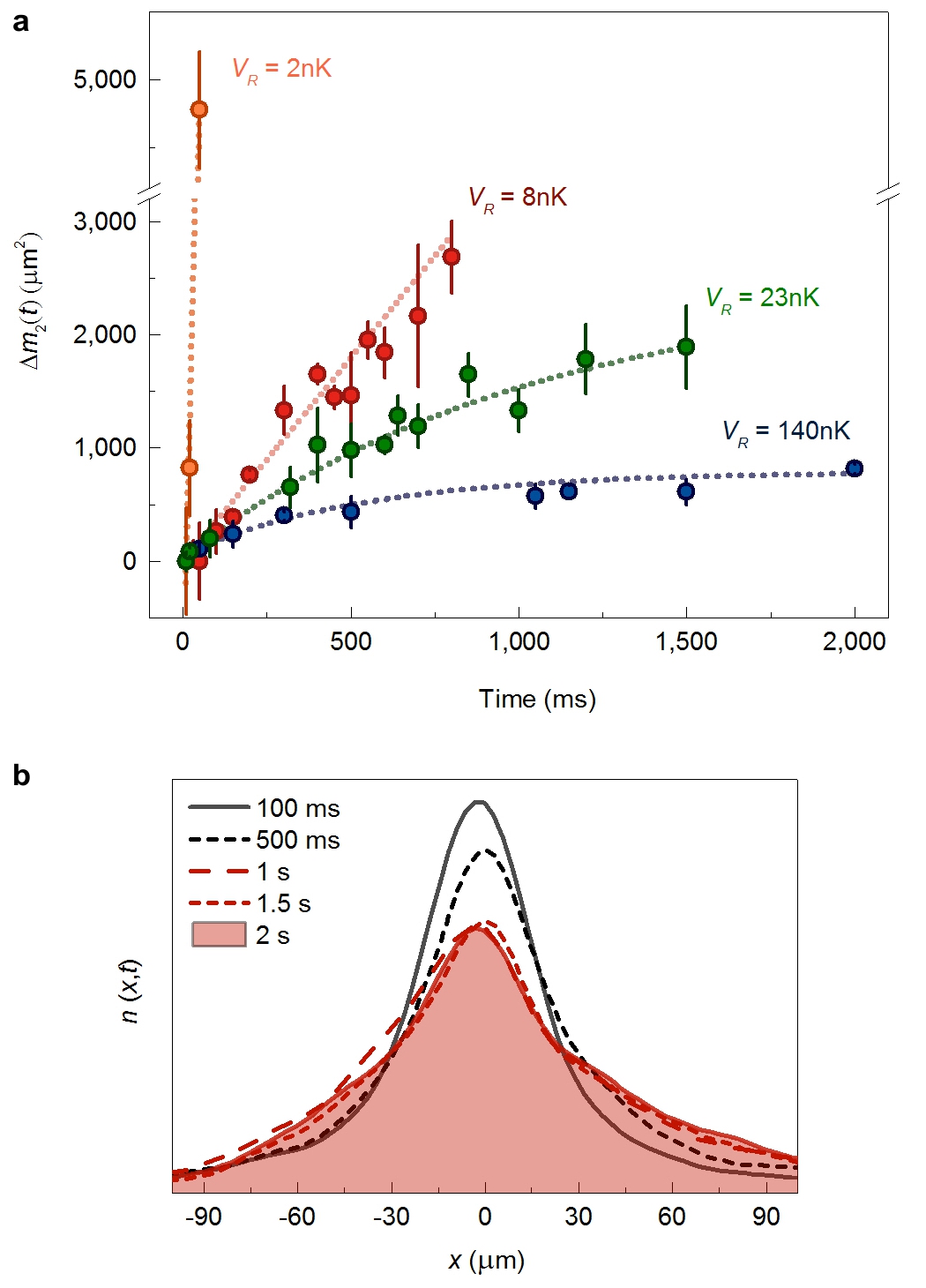}
\label{fig2}
\caption{\textbf{Expansion and localization dynamics}. \textbf{a}) Time evolution of the second moment of $n(x)$ for different disorder strengths. Three typical regimes are reported: diffusion (orange and red), anomalous diffusion (green) and localization (blue). \textbf{b}) Time evolution of the density for $V_R$=46.8(1.4)nK.}
\end{figure}

To prepare a low-energy system in the disorder, we slowly increase $V_R$ from zero to a finite value while reducing to zero both the interactions and the harmonic confinement. In order to characterize the diffusive or localized nature of the system, we then let it evolve in the disordered potential for a variable time and we finally image the spatial distribution using absorption imaging. We study in particular the one-dimensional density distribution $n(x)$, obtained from the integration along two spatial dimensions (an example of the corresponding evolution along $y$ and $z$ is reported in Extended Data Fig.3). Measuring the integrated second moment $m_2$ of $n(x)$, we observe a strong dependence of the expansion dynamics on $V_R$ (Fig.2a). For few smaller values of $V_R$, the evolution is purely diffusive, according to $\Delta m_2(t)=m_2(t)-m_2(t=0)=2Dt$. The linear increase of $m_2$ turns into anomalous diffusion for intermediate disorder and eventually, for larger $V_R$, a small initial increase of $m_2$ is followed by a plateau, indicating that any further expansion of the cloud is inhibited and the system is fully localized. Fig.2b shows a typical time evolution of $n(x)$ in the large-$V_R$ regime; there is an initial expansion of the tails that explains the short-time increase of $m_2$, while for longer times the shape of $n(x)$ apparently stops changing.

\begin{figure}
\includegraphics{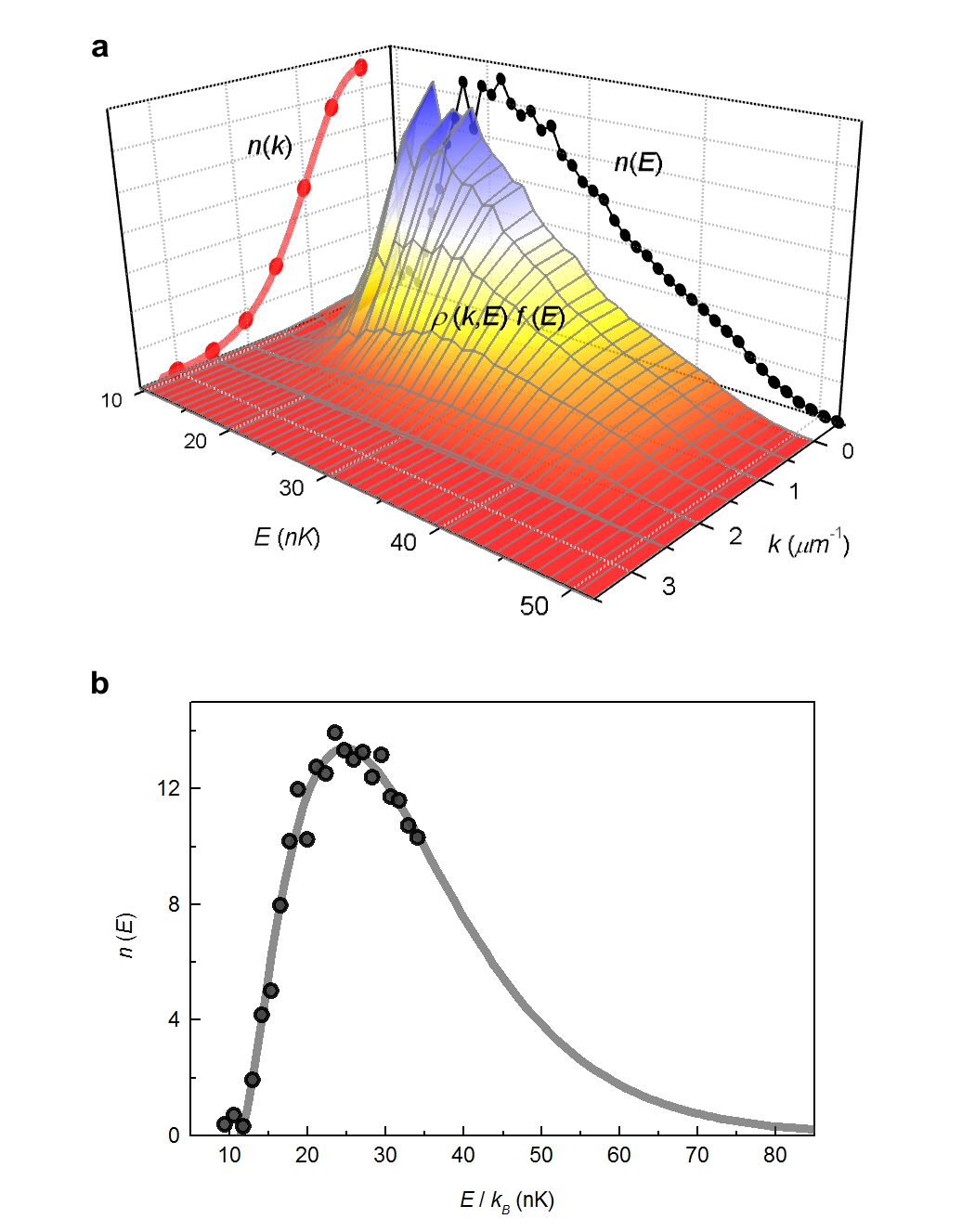}
\label{fig3}
\caption{\textbf{Momentum and energy distribution}. \textbf{a}) Sketch of the reconstruction process of $n(E)$ from the measured $n(k)$ and the calculated $\rho(E,k)$. \textbf{b}) Reconstructed energy distribution. The data are limited to the energy range where the finite-size simulations are considered reliable and fitted with $n(E)=(E-E_{0})^\alpha \exp((E-E_{0})/E_m)$ (see Extended Data Fig.7 for further details).}
\end{figure}

Studying the short-time evolution of the cloud, we also find that the breakdown of the diffusive transport for $V_R/k_B>$10nK is associated to an abrupt reduction of the short-time diffusion coefficient, down to $D\approx\hbar/3m$. This is indeed the predicted transport regime where quantum interference should suppress diffusion and lead to localization (see Extended Data Fig.5).

\begin{figure}
\includegraphics{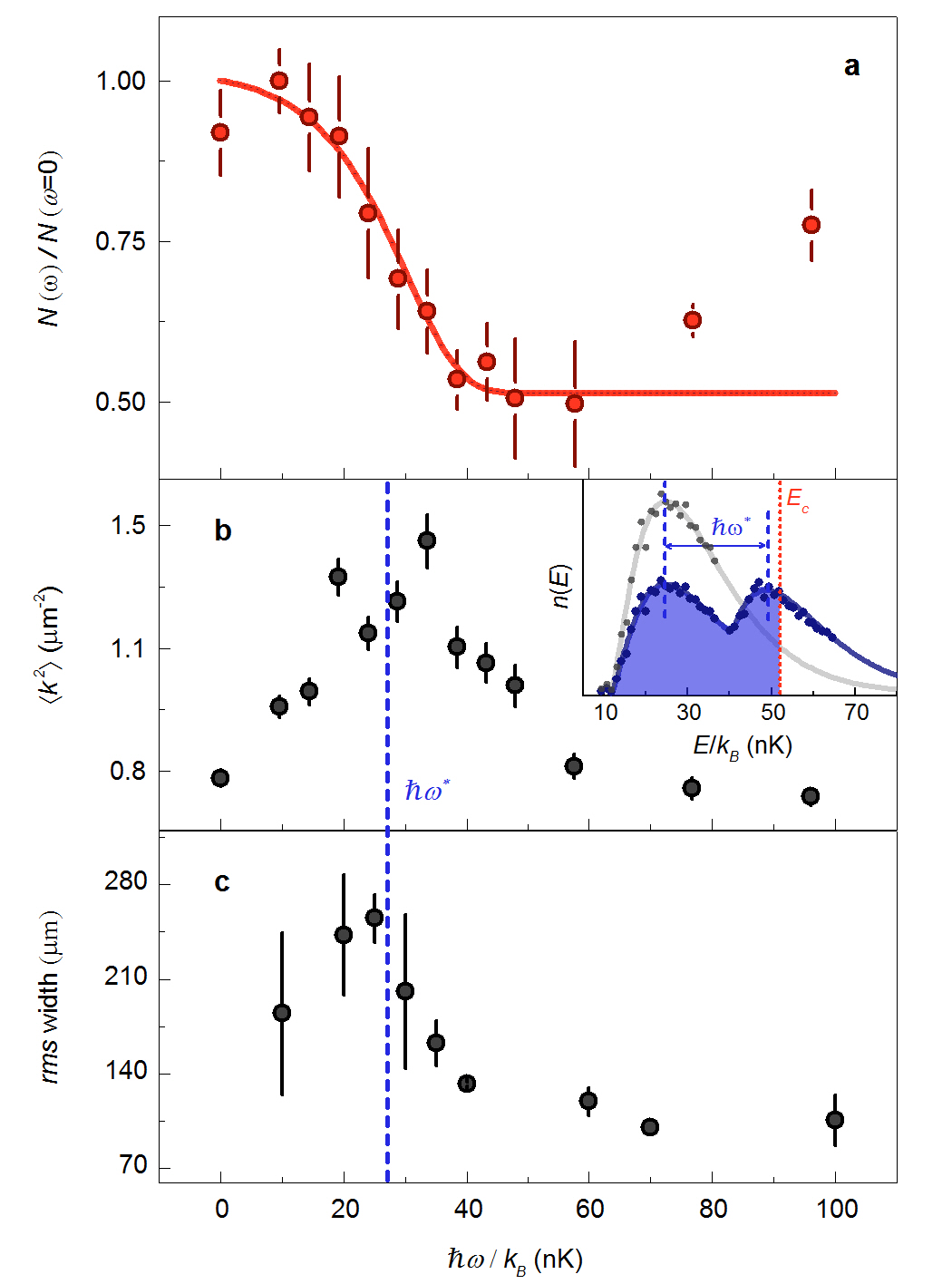}
\label{fig4}
\caption{\textbf{Excitation spectrum}. Measured evolution of the atom number (\textbf{a}), of the kinetic energy (\textbf{b}) and of the spatial size (\textbf{c}) vs the modulation frequency, for $V_R$=46.8(1.4)nK. A fit of the atom number with the excitation model described in the text (continuous line) gives the mobility edge at $E_c$=52(5)nK. Inset: $n_0(E)$ (grey dots) and the calculated $n(E)$ (blue dots) after modulating at $\omega^{*}$; the position of $E_c$ is obtained from the fit in \textbf{a}).}
\end{figure}

The observation of a rather sharp transition to localization with increasing $V_R$ suggests that the energy spread of the system is narrow. We can experimentally access only the distribution of kinetic energy, which is measured through a standard time-of-flight technique. Just after loading the sample into the speckles, we release it abruptly from the disordered potential and we reconstruct the one-dimensional momentum distribution $n(k)$ along the $x$ direction from the time evolution of the density (see Extended Data Fig.6). We find that in the localized regime, i.e. for $V_R/k_B>$ 18nK, the mean kinetic energy always lies well below $V_R$. This motivated us to perform a numerical study of the low-energy eigenstates by exact diagonalization of the disorder Hamiltonian, to reconstruct the total energy distribution from the measured $n(k)$. Numerical simulations are notoriously hard for energies close to the mobility edge, due to the finite spatial size in the simulations, but we found that a box with side length of about 10$\pi\sigma_R$ could give reliable results in the low-energy range occupied by the system. We calculate the momentum-space density of the energy eigenstates $\rho(E,k)$, which allows to relate $n(k)$ to the energy distribution function $f(E)$ through $\int \rho(E,k) f(E) dE=n(k)$. In practice, we search for the $f(E)$ that best reproduces the experimental $n(k)$, as sketched in Fig.3. We find that an exponential form, $f(E)\propto\exp(-E/E_m)$, provides a very good agreement with the experimental data. The energy distribution is then determined as usual as $n(E)=\int \rho(E,k) f(E) dk$. The typical $n(E)$ is narrow and peaked at an energy not far from the lowest energy $E_{0}$ (Fig.3b), in agreement with the observation of fully localized systems.

The final step to determine the mobility edge, from now on $E_c$, consists in producing a controlled energy excitation in the system, so as to promote the initial $n(E)$ towards diffusive high energy states. After loading the atoms into the disorder, we apply a weak sinusoidal modulation to the laser power for 0.5s. This corresponds to a time-dependent perturbation of the disordered potential $V_R(t)=V_R(1+A\cos(\omega t))$, with $A\approx0.2$ and variable $\omega$. In the small $A$ limit this procedure allows to excite a fraction of the atoms by exactly $\Delta E=\hbar\omega$. The final energy distribution can be written as $n(E,\hbar\omega)=(1-p)n_0(E)+p n_0(E-\hbar\omega)$, where $n_0(E)$ is the initial energy distribution and $p=p(E,\omega)$ is the probability to excite an atom at energy $E$ to $E+\hbar\omega$. Despite $p$ is in principle depending both on $E$ and $\omega$, in the Supplementary Information we estimate that this dependence is weak in the relevant range of energies, so that we can take $p$ as a constant. At the end of the modulation sequence we leave the disorder at fixed $V_R$ for another 0.5s, allowing the atoms transferred to diffusive states to expand enough to be effectively not visible to our imaging system. The transfer to diffusive states is hence detected as atom losses.

Fig.4 shows an example of the final atom number measured for different modulation frequencies. To determine $E_c$ we fit the data $N(\omega)$ with $N_{loc}(E_c,\hbar\omega)=\int^{E_c}_0{n(E,\hbar\omega) dE}$, where $E_c$ is the only free parameter. For the specific dataset in Fig.4 we obtain $E_c=52(5)$nK. The agreement between the data and the model is in general very good until a large-$\omega$ regime where an unexpected increase of $N$ occurs. This can be justified considering the poor overlap in momentum space between deeply localized states and the essentially free states of the continuum, which reduces the excitation probability $p$. In absence of a precise modelling, we exclude the data at high frequency from the fit. A test of the validity of our model is provided by the evolution of both system size and kinetic energy with $\omega$ (Fig.4b-c). They indeed reach a maximum value around the same frequency $\omega^{*}$ for which the model indicates an optimal transfer to localized states just below $E_c$, which therefore have the largest localization lengths and the largest energies (see the computed $n(E,\hbar\omega^{*})$ in the inset of Fig.4b). For $\omega>\omega^{*}$, the peak of the excited part of $n(E)$ crosses $E_c$ and an increasingly larger fraction of atoms diffuses away and gets effectively lost, not contributing to the size nor to the kinetic energy, which therefore decrease again.

We repeated this procedure for several disorder strengths; a summary of the measured trajectory for $E_c$ in the disorder-energy plane is shown in Fig.5. The most interesting regime for Anderson localization is the one where $V_R<E_R$, since trapped states in individual wells of the disorder are extremely rare, and localization is caused by destructive interference on the single-particle wavefunction over many wells and barriers. In this regime we observe an almost linear scaling of $E_c$ with $V_R$, which is justified by the fact that $V_R$ is the only relevant energy scale in the system. When $V_R$ is increased above $E_R$, the trajectory bends down. A saturation of $E_c$ for large $V_R$ might actually be expected. When $V_R\gg E_R$, so that the tunneling through the random barriers is suppressed, the particle mean free path reaches a minimum value which is set by the mean distance between the barriers, i.e. $\sigma_R$. According to the Ioffe-Regel criterion \cite{Ioffe}, this corresponds to a maximum kinetic energy at the transition.

\begin{figure}
\includegraphics{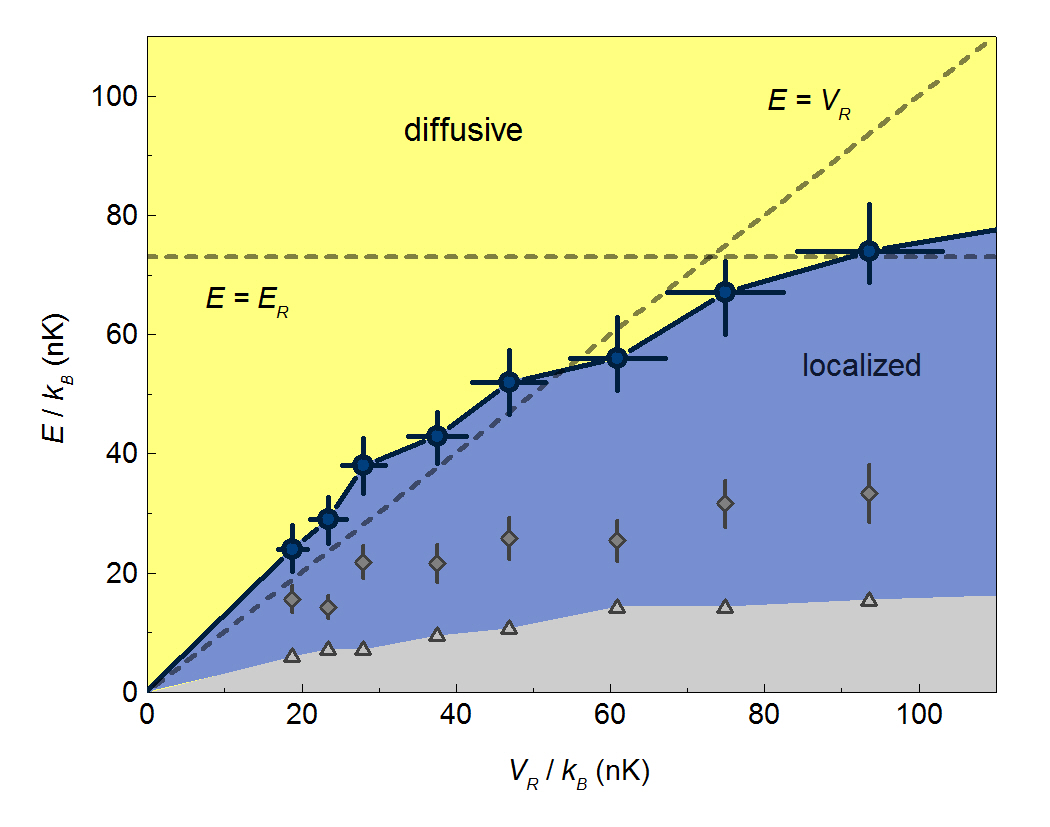}
\label{fig5}
\caption{\textbf{Mobility edge vs disorder strength}. Trajectory of $E_c$ (blue circles) separating localized (blue) from diffusive (yellow) states. Grey diamonds mark the position of the peaks of $n_0(E)$, while grey triangles stand for the lowest energy $E_0$. The vertical error bars for $E_c$ contain both uncertainties in the determination of $n_0(E)$ and in the fit of $N(\omega)$, while the horizontal ones represent the uncertainty in the determination of $V_R$ (details on the calibration of $V_R$ are reported in the Extended Data Fig.4).}
\end{figure}

Our results are rather different from a previous experimental report of $E_c\approx 2V_R$, relative to a speckle potential with different spatial correlations \cite{Kondov}. That experiment was however performed on a much shorter time scale and the analysis neglected the potential energy of the disorder, both approaches which do not allow a direct comparison with the present results. We also find only a qualitative agreement with theoretical determinations of $E_c$ for isotropic speckles \cite{Yedjour,Piraud2,Delande}. In particular, our results for $E_c$ are about 20\% higher than those based on self-consistent approaches \cite{Yedjour,Piraud2}, or about 40\% higher than those obtained via numerical computation \cite{Delande}. These deviations might be at least partially attributed to the different anisotropy of the speckles in experiment and theory, and are definitely worth further studies. We note that a comparison to lattice models \cite{Anderson} could be carried out only close to the band edge, where the lattice contributes only with an effective mass, but obtaining indisputable results in that regime has proven to be challenging \cite{Bulka}.

In conclusion, we have experimentally determined the mobility edge trajectory in a system with controlled microscopic disorder and tunable energy. This will allow for the first time a direct comparison of an experiment to numerical calculations and may also provide a test of analytical theories. A further narrowing of the energy distribution could allow us to measure in the future also the critical properties at the transition. Our technique is general for atoms and can be applied to other types of disorder. A full assessment of the non-interacting problem is the prerequisite for exploring challenging problems for interacting disordered systems, such as many-body localization \cite{Basko}, anomalous diffusion \cite{Cherroret}, or Bose-Einstein condensation \cite{Huang,Natterman,Pilati}, where already making theoretical predictions becomes very hard.

%\begin{methods}
%We might put just a very concise Methods Summary, and then write a longer Supplementary Information.
%
%\end{methods}

%% Put the bibliography here, most people will use BiBTeX in
%% which case the environment below should be replaced with
%% the \bibliography{} command.

\vspace{1cm}

\small
\begin{itemize}
	\item[] *These two authors contributed equally. 
	\item[] \textbf{Acknowledgements} We acknowledge discussions with V. Josse, L. Pezz\'e and D. S. Wiersma. This work was supported by ERC (grants 247371 and 258325).
	\item[] \textbf{Author Contributions} G.S and M.L. designed the experiment; G.S, M.L. and G.M. analysed the data and performed the numerical simulations; all the other authors participated to the experiment, data analysis, discussion of the results and writing of the manuscript.
	\item[] \textbf{Correspondence} Correspondence and requests for materials should be addressed to G.M.~(email: modugno@lens.unifi.it).
\end{itemize}

\clearpage

\onecolumngrid
\renewcommand\thefigure{ED\arabic{figure}}
\setcounter{figure}{0} 
\begin{center}
\textbf{\large EXTENDED DATA FIGURES}
\end{center}
\vspace{2cm}
\begin{figure*}[htbp]
\includegraphics{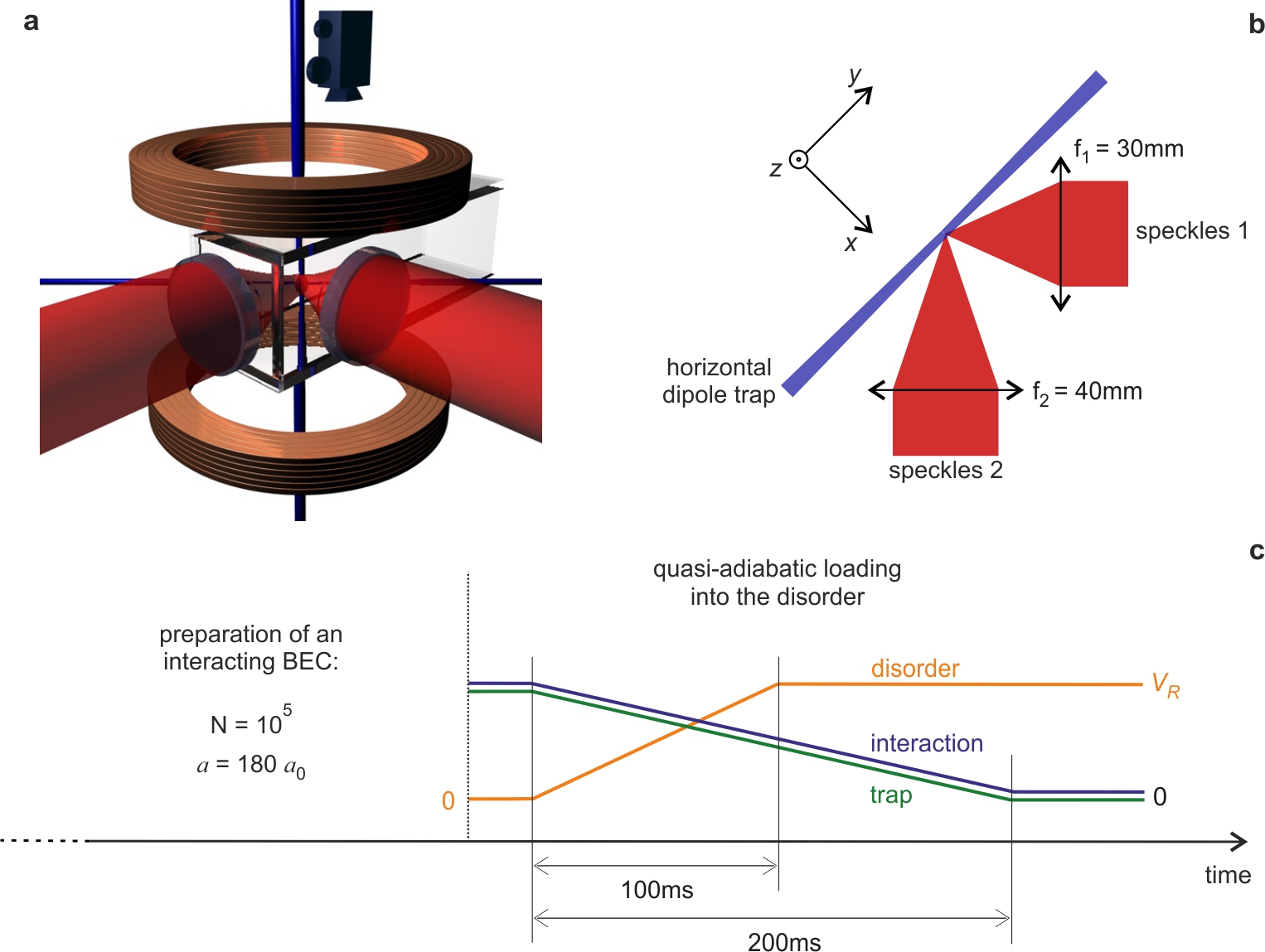}
\caption{\textbf{System geometry and preparation.} \textbf{a,b}) The dipole trap we initially use to prepare the Bose-Einstein condensate is the result of two crossed 1064nm laser beams (blue), providing trap frequencies of 110Hz along $x$ and $z$ and 25Hz along $y$. The disordered potential is created by the interference of two focused speckles beams, intersecting at 90$^\circ$ at the atoms position (red). The speckles wavelength is $\lambda$=762nm, blue-detuned with respect to the $^{39}$K optical transitions (767nm and 770nm). We use two pairs of coils to tune the scattering length $a$ via a magnetic Feshbach resonance and to compensate gravity (for more details, see Ref.[32]). We have two available imaging systems: one on the $z$ axis and the other one along the same direction of the speckles 1. This allows us to study the expansion dynamics of the cloud along all the three main axes. We normally use the imaging along $z$ (sketched here), since we are especially interested in the dynamics along $x$. \textbf{c}) We realize a quasi-adiabatic loading of the atoms into the disorder by slowly ramping down the trap and the interactions while rising up the speckles: the procedure was optimized by minimizing both size and kinetic energy of the system localized in relatively strong disorder ($V_R>$50nK).}
\end{figure*}

\begin{figure*}[htbp]
\includegraphics{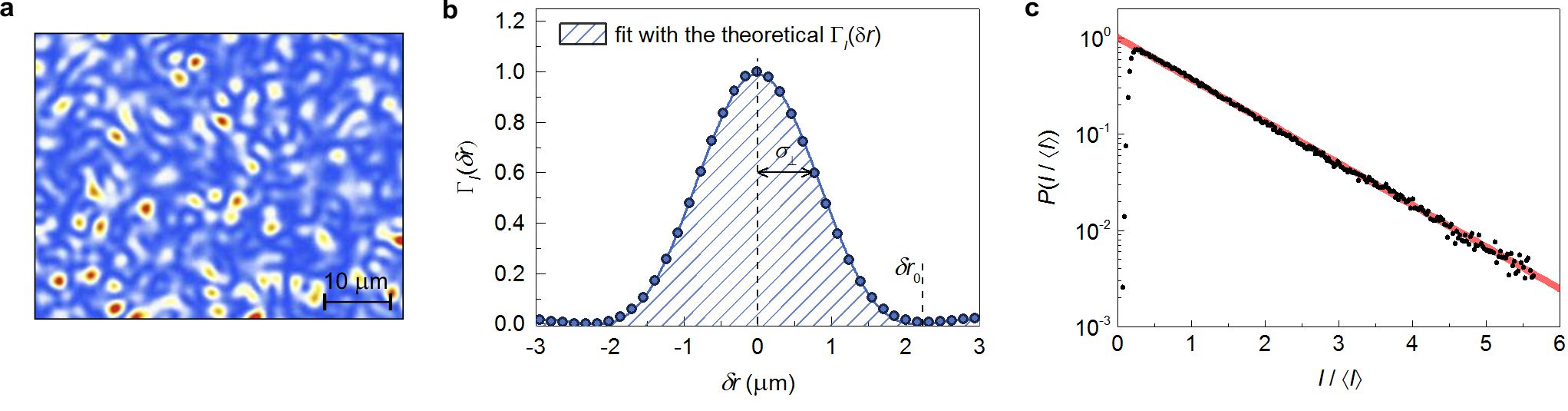}
\caption{\textbf{Characterization of the disordered potential.} We image ex-situ the two speckles beams at their focus position with a CCD camera and we measure the waists of the Gaussian envelopes ($s_1$=1280$\mu$m and $s_2$=1315$\mu$m). \textbf{a}) Using a 35x microscope we can detect the fine structure of the speckles. (\textbf{b}) From these images we can derive the intensity autocorrelation function $\Gamma_{I}(\delta r)$, which is well fitted by the theoretical Bessel function profile (see Ref.[31]). The transverse correlation length is given by $\sigma_{\bot}=\delta r_0/\pi$, where $\delta r_0$ is the position of the first zero. We measure $\sigma_{\bot 1}$=0.73$\mu$m and $\sigma_{\bot 2}$=0.80$\mu$m. \textbf{c}) For each speckles we also derive the intensity distribution function $P(I)$. We find that it decays exponentially according to $P(I)=1/\langle I\rangle \exp(-I/\langle I\rangle)$, as expected for fully developed speckles (Ref.[31]). The deviation at low $I$ is due to the finite resolution of the imaging system used to calibrate the speckles.}
\end{figure*}

\begin{figure*}[htbp]
\includegraphics{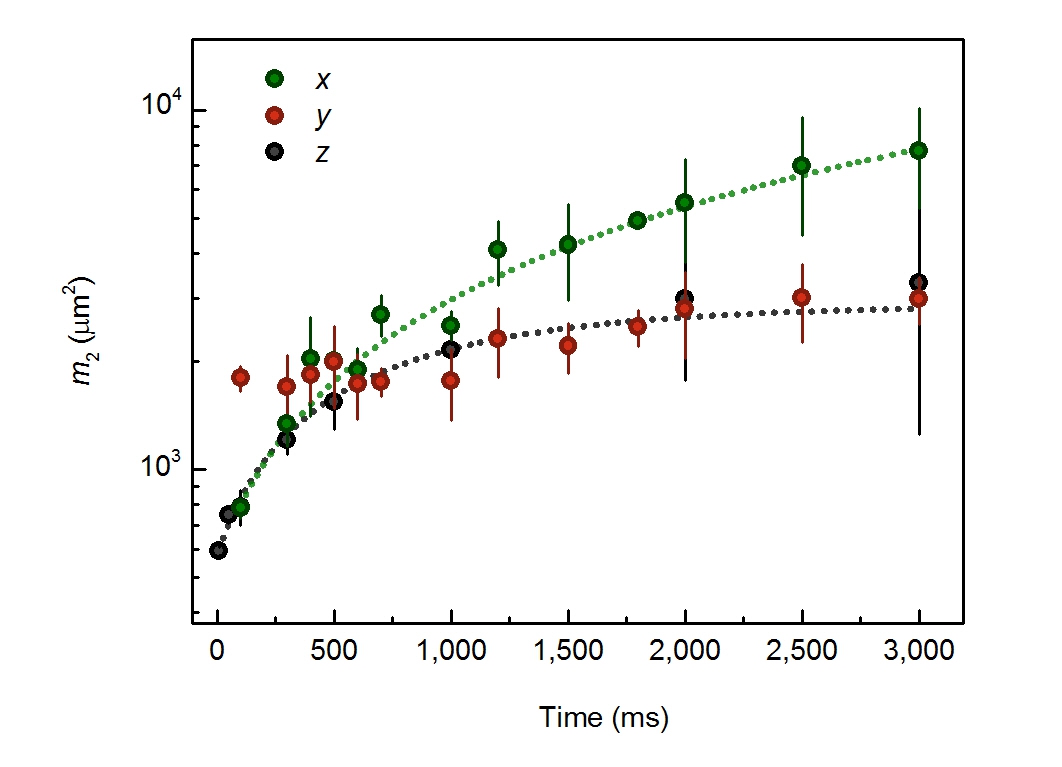}
\caption{\textbf{Expansion in the anisotropic disorder.} To estimate the effect of the anisotropic correlation function of the disorder, we look at the time evolution of the squared width of the spatial distribution $n(\textbf{r})$ along the three main axes, for an intermediate value of the disorder amplitude $V_R$=23.4nK. At short times the system expands along both $x$ and $z$, eventually reaching a larger size along $x$, where the disorder correlation length is a factor $\sqrt{2}$ larger than along $z$. Along $y$ we observe just a slight increase of $m_2$. This might be consistent with the larger initial size, due to the anisotropy of the dipole trap, and the smaller correlation length caused by interference. In order to have the highest sensitivity to changes in the diffusion/localization dynamics, in our quantitative analysis we consider the evolution along $x$.}
\end{figure*}

\begin{figure*}[htbp]
\includegraphics{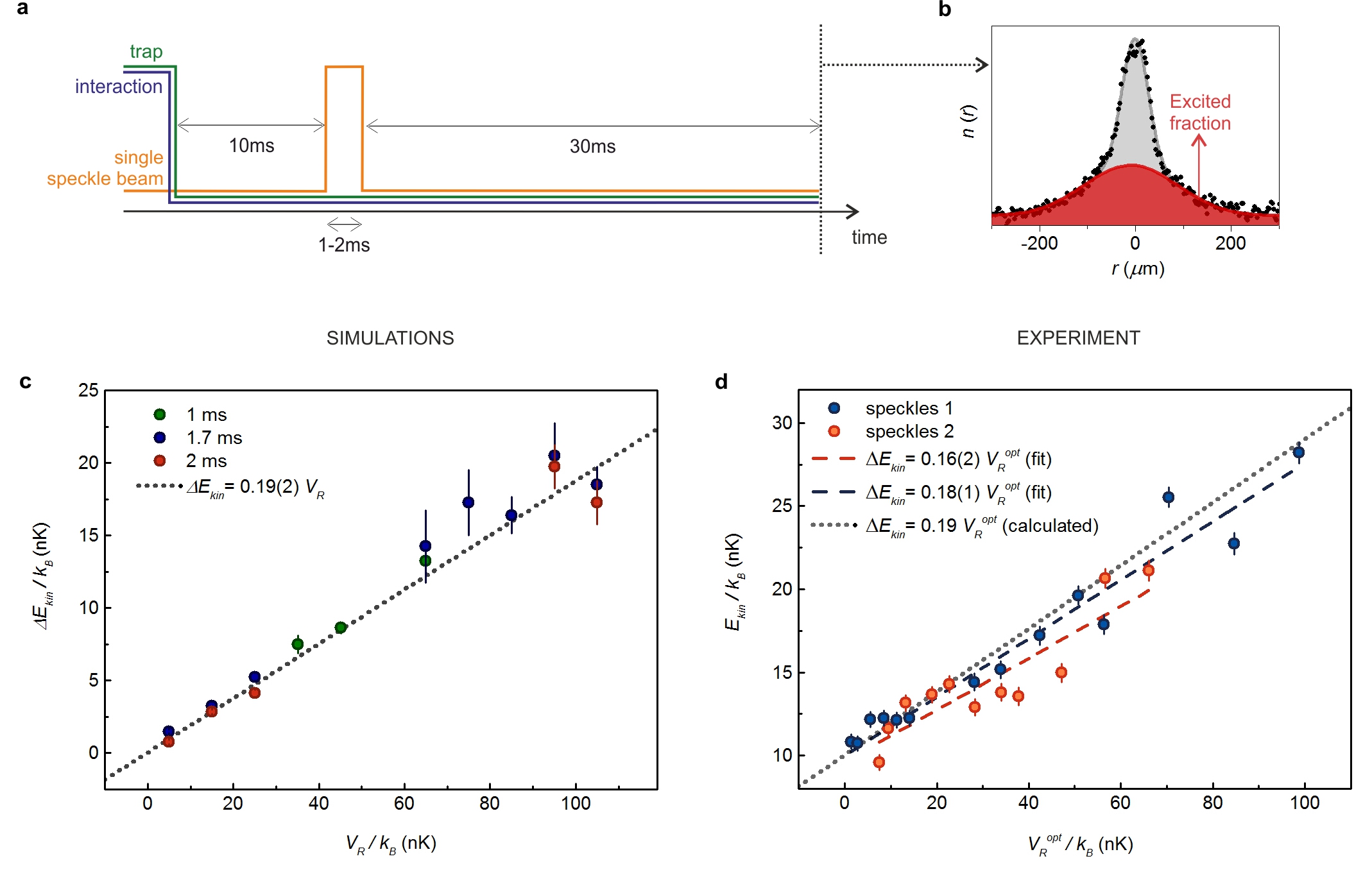}
\caption{\textbf{Calibration of the disorder amplitude.} $V_R$ is calibrated using two different methods. First, we compute a $V_R^{opt}$ from the light intensity of the speckles, obtained by measuring ex-situ the spatial envelope of the individual speckle beams and their total optical power. For this optical calibration we estimate a relative uncertainty of 10\%. Second, we perform an in-situ calibration from the dynamics of the atoms. Since our 3D disorder is the superposition of two separate speckle patterns, we calibrate them independently. \textbf{a}) We apply a short pulse of the speckle potential to a non-interacting condensate in free expansion, and we measure its final momentum distribution. Using a single speckle beam at a time, the problem is effectively 2D, being the longitudinal correlation length much longer than the atomic displacement on the short-time scale we consider ($\sigma_{\parallel}\simeq 10 \sigma_{\bot}$). \textbf{b}) We observe the formation of a bimodal momentum distribution, meaning that only a fraction of the atoms gets accelerated by the pulse. \textbf{c}) From a 2D numerical simulation we indeed observe that the momentum distribution should develop a high-energy component, with a mean kinetic energy transferred by the speckles pulse $\Delta E_{kin}=0.19(2) V_R$. \textbf{d}) The behavior we observe in the experiment is in agreement with the simulations. The evolution of $E_{kin}$ of the excited part is approximately linear with $V_R^{opt}$, with a slope within 17\% of the theoretical one. This confirms the validity of the optical calibration. The disorder strength reported in the paper is that obtained with the optical calibration, $V_R=V_R^{opt}$.}
\end{figure*}

\begin{figure*}[htbp]
\includegraphics{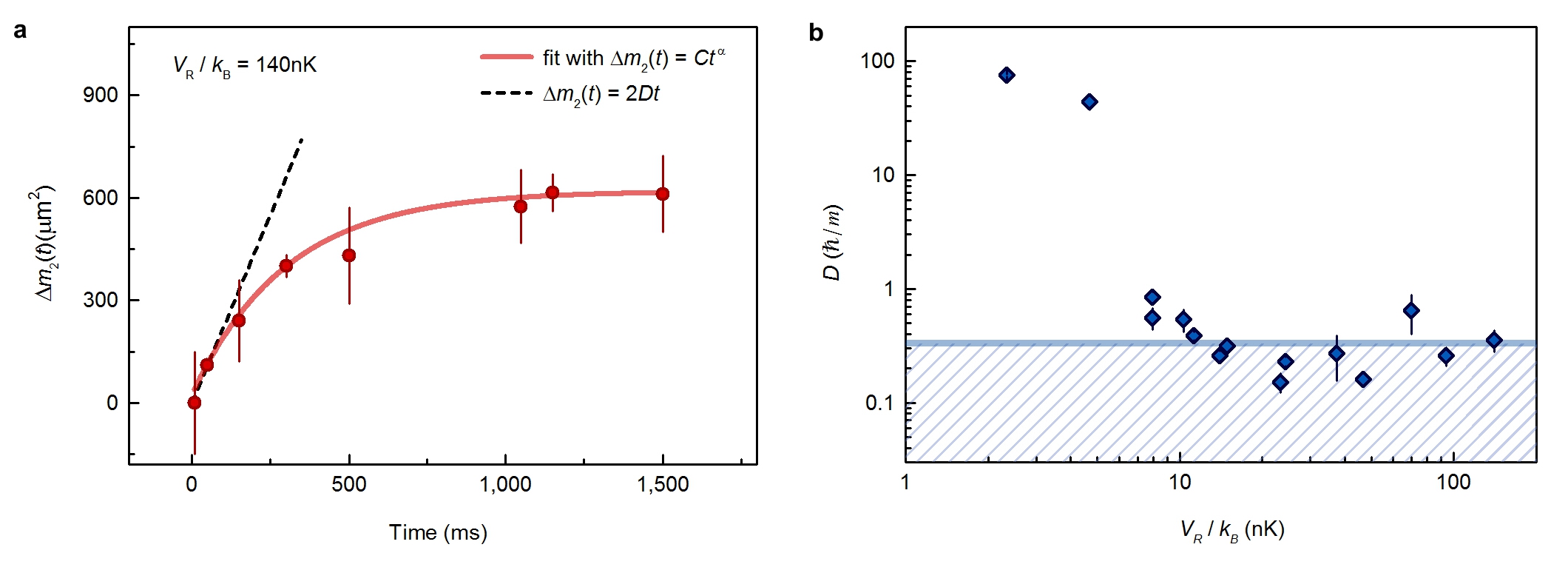}
\caption{\textbf{Diffusion constant measured at short time vs $V_R$.} In Fig.2 in the main paper we observed that, in the localized regime, the density profile always undergoes an initial expansion before reaching its equilibrium size. We study the dynamics of this initial evolution and we compare it to the diffusive dynamics for lower disorder. \textbf{a}) To do this we fit the data with $\Delta m_2(t)=Ct^a$ and we estimate an initial diffusion coefficient as $D=dm_2/2dt$ at $t=0$. \textbf{b}) We observe a rapid drop of $D$ for increasing $V_R$ followed by a saturation around $\hbar/3m$ for $V_R/k_B>$ 10nK. According to perturbative self-consistent theories (Ref.[33]), the diffusion coefficient in a 3D disordered system is $D=\hbar kl/3m$, where $l$ is the mean free path and $k$ is the atomic wavevector. The Ioffe-Regel criterion predicts the onset of localization for $kl\simeq$1, i.e. $D\simeq\hbar/3m$, which is indeed the typical value we measure for $V_R>$10nK. This means that we observe partial or complete localization - indicated by the bending and eventual flattening of $m_2(t)$ (Fig.2) - when the system enters the quantum transport regime (shaded area). This is indeed the regime where interference effects - at the basis of Anderson localization - come into play.}
\end{figure*}

\begin{figure*}[htbp]
\includegraphics{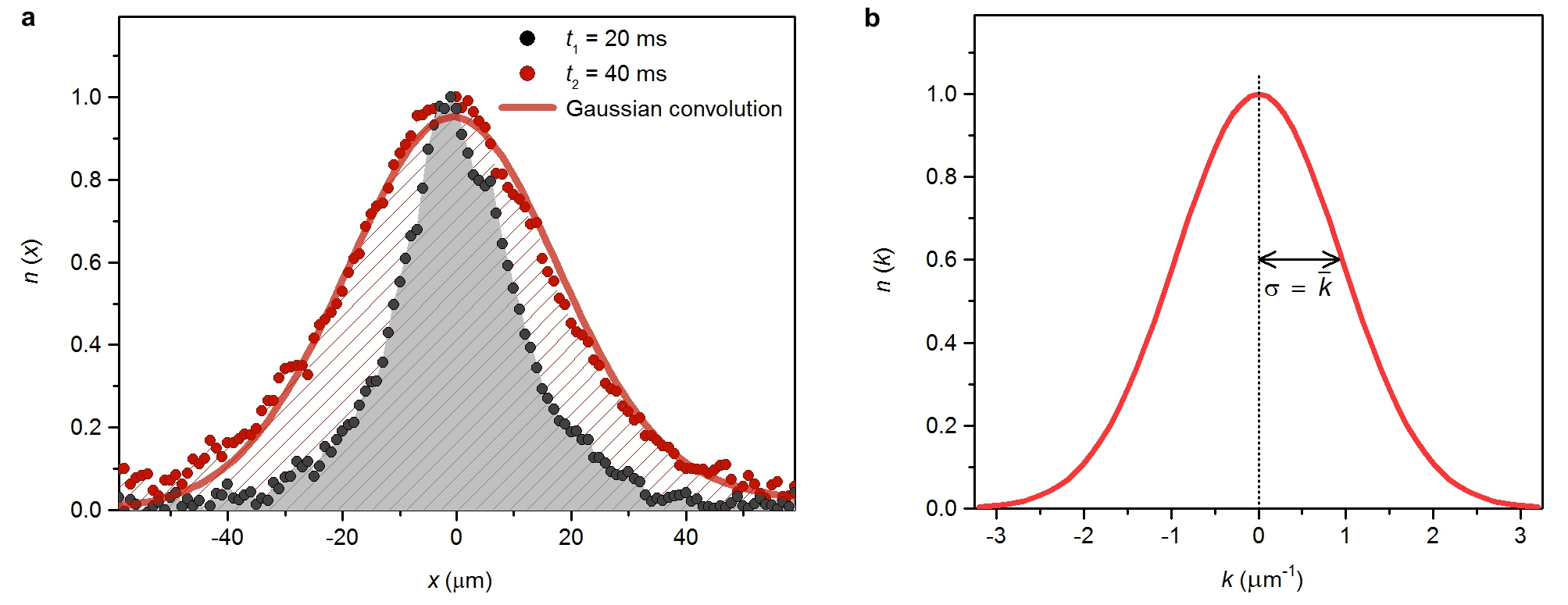}
\caption{\textbf{Momentum distribution.} \textbf{a}) After loading the atoms into the disorder, we switch off the speckles and we measure the atomic density at two different times of flight, $t_1$ and $t_2$ (we report here an example for $V_R/k_B$=46.8(1.4)nK). We then use a deconvolution procedure to deduce the momentum distribution. The spatial density at time $t_2$, integrated along $y$ and $z$, is given by $n(x,t_2)=\int{dk dx_1 n(x_1,t_1) n(k) \delta(x-x_1-\frac{\hbar k}{m}(t_2-t_1))}$, where $k=k_x$ and $n(k)$ is the momentum distribution integrated along $k_y$ and $k_z$. Here we have assumed the momentum and spatial distributions to be factorizable. The system indeed occupies a large number of states and we can reasonably assume that there are no relevant correlations in the average distributions. We find that a Gaussian form of $n(k)$ well reproduces the data; then we substitute $n(k)\propto\exp (-k^2/(2\sigma^2))$ in the previous formula and we get $n(x,t_2)=\int{dx_1 n(x_1,t_1) \exp{\left[-\frac{1}{2}\left(\frac{m(x-x_1)}{\sigma (t_2-t_1)}\right)^2\right]}}$. We include the experimental $n(x_1,t_1)$ in the integral and we use it to fit the density distribution at time $t_2$, with $\sigma$ as the only free parameter. \textbf{b}) We then extract $n(k)$ and we use it to deduce $n(E)$ as in Fig.3a in the main paper. From the value of $\sigma=\bar{k}$ we can calculate the mean kinetic energy as $E_{kin}=3(\hbar^2 \bar{k}^2/2m)$, which for this dataset is 16.5nK.}
\end{figure*}

\begin{figure*}[htbp]
\includegraphics{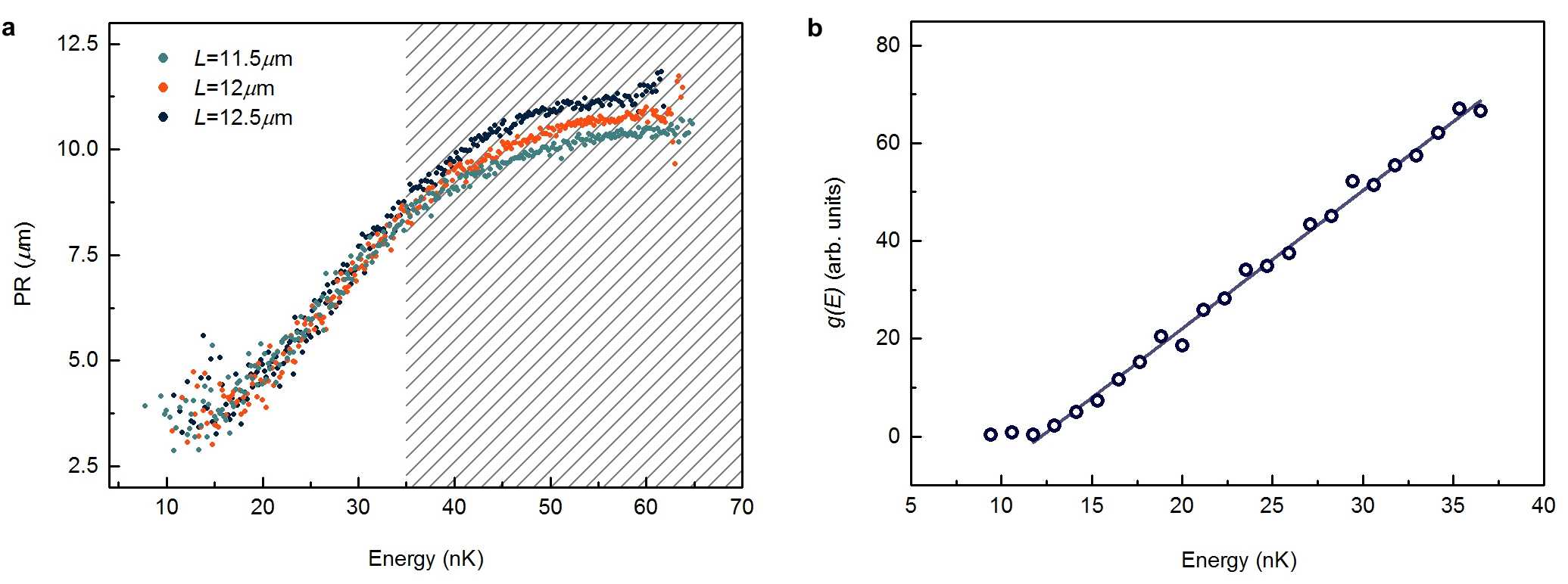}
\caption{\textbf{Numerical study of low energy eigenstates.} We solve the single-particle Schr\"{o}dinger equation by exact diagonalization of small size systems. We employ a synthetic 3D potential generated by the interference of two crossed speckles with transverse correlation length $\pi\sigma_R$=2.4$\mu$m, corresponding to the average of the experimental ones. We have neglected the longitudinal evolution of each speckle field, which is on a much larger lengthscale. The typical system is a cube with side length $L$=12.5$\mu$m and a discretization length of 0.25$\mu$m. The axes of the cube are along the three principal axes of the crossed speckle field. All results are averaged over at least 50 different realizations of the disorder. \textbf{a}) For each eigenstate we evaluate the one-dimensional participation ratio $1/\int(\int\psi(x,y,z)^2dydz)^2dx$, which is a measure of the characteristic length of the wavefunction integrated along two spatial directions. The evolution of the participation ratio with the energy and $L$ indicates that the results are not affected by the finite system size up to an energy typically around 0.75$V_R$. In the example reported here for $V_R$=46.8nK the curves corresponding to three different $L$ are indistinguishable for $E\leq$35nK, and start to deviate only for larger energy. \textbf{b}) The average density of states $g(E)=\int \rho(E,k) dk$ extracted from the simulations shows a power law scaling for energies larger than a certain $E_0$; below $E_0$, $g(E)$ is very small and drops rapidly to zero. The typical exponents for $g(E)$ are between 1 and 2, and they grow with $V_R$: in the example reported here for $V_R$=46.8nK, we find $\alpha\simeq$1. According to this scaling of $g(E)$, we use the function $n(E)=g(E) f(E)=(E-E_0)^\alpha \exp(-(E-E_0)/E_m)$ to fit the data of $n(E)$ obtained as in Fig.3 in the paper, where $E_m$ is fixed by the reconstruction process (Fig.3a) while $E_0$ and $\alpha$ are free parameters.}
\end{figure*}

\begin{figure*}[htbp]
\includegraphics{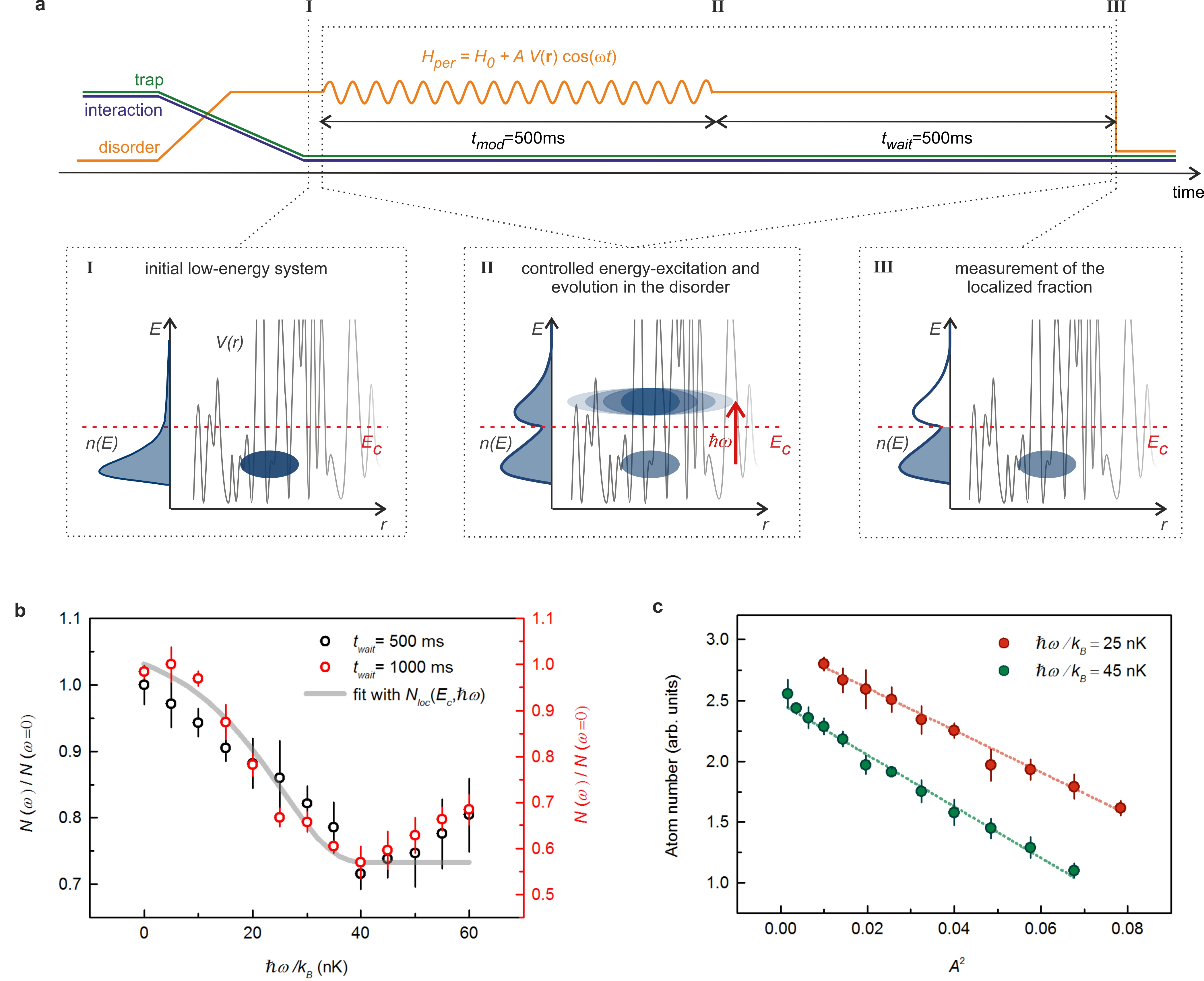}
\caption{\textbf{Modulation spectroscopy.} \textbf{a}) Scheme of the experimental sequence. \textbf{b}) We choose the duration of the modulation and the successive waiting time according to the typical time scales we observe in the $m_2$ measurements: $t_{mod}+t_{wait}$=1s. To be sure that after this time interval the diffusive fraction has expanded enough to be no longer detectable, we have measured the evolution of the loss spectrum with $t_{wait}$. Here we show the losses at $t_{wait}=$0.5s and 1s, for $t_{mod}$=0.5s and $V_R$=46.8(1.4)nK. The losses for $t_{wait}$=1s are slightly larger, but the global behavior is the same. The two spectra indeed provide the same estimation for $E_c$. This indicates that $t_{wait}$=500ms used in the experiment is enough to detect the transfer of atoms above $E_c$. \textbf{c}) Our model of the excitation process is built in the framework of the Fermi's golden rule: given the perturbed Hamiltonian $H_{per}=H_0+A V(\textbf{r}) \cos(\omega t)$, the excitation probability is $p(E,\omega)=A^2 \sum_{i,f}{\left|\langle f\left|V(\textbf{r})\right|i\rangle\right|^2 \delta(E_i-E) \delta(E_f-(E+\hbar\omega))}$, where $E$ is the initial energy and $\omega$ is the modulation frequency. To test the validity of this perturbative approach, we verify that the atom number at the end of the excitation sequence, which is proportional to 1$-p(E,\omega)$, scales linearly with $A^2$. We repeat the measurement at two different $\omega$ for $V_R$=46.8(1.4)nK and we indeed observe a linear scaling with $A^2$, confirming the validity of our approach.}
\end{figure*}

\clearpage
\begin{center}
\textbf{\large SUPPLEMENTARY INFORMATION}
\end{center}
\vspace{1cm}
\twocolumngrid

\section{Characterization of the 3D disordered potential}
In Extended Data Fig.2, we have fully characterized the two speckles beams generating the disordered potential. We now use those measurements to reconstruct the corresponding properties of the final 3D potential. When the beams cross at 90$^\circ$ and the interference pattern adds to the speckles modulation, the main axes of the problem are the ones in Fig.1. In the $y$ direction, interference fringes split the speckles into smaller substructures and the first zero in the correlation function $\delta r_0$ is given by half the distance between two interference maxima, so that $\sigma_{y}=\lambda/(2 \sqrt{2}\pi)$=0.09$\mu$m. Along $z$ we calculate the average between the transverse correlation lengths of the two speckles $\sigma_{z}=(\sigma_{\bot 1}+\sigma_{\bot 2})/2$=0.76$\mu$m, while $\sigma_x$ is the projection at 45$^\circ$ of the same average: $\sigma_{x}=\sigma_{z} \sqrt{2}$=1.08$\mu$m. From the geometric average on the three directions we get $\sigma_R$=0.41$\mu$m.

The other main feature of the disordered potential is the intensity distribution $P(I)$. The field generated by two interfering speckles is predicted to preserve the same exponential distribution of the single speckles with $\langle I_{tot}\rangle=\langle I_{1}\rangle+\langle I_{2}\rangle$ (see Ref.[31]). An angle $\Delta\theta$ between the polarizations of the two speckles would reduce the contrast in the interference and hence modify $P(I)$ at low $I$. Considering the geometry of our experimental setup, we estimate $\Delta\theta\leq5^\circ$. For small $\Delta\theta$, the position of the maximum in $P(I)$ is expected to move to $\Delta\theta^2/4 \ln(4/\Delta\theta^2) \langle I_{tot}\rangle$, which in our case corresponds to $\approx 0,01 \langle I_{tot}\rangle$. Such a small modification is not expected to affect localization properties in the system, since the typical energies in the experiment range from $V_R/2$ to $V_R$, hence far from this low $I$ region.

\section{Background potential}
In addition to the disordered speckle potential, the atoms are subjected to a weak additional optical and magnetic potential. We indeed use a vertically-oriented homogeneous magnetic field to control the interaction, while a magnetic field gradient compensates gravity. The two sets of magnetic field coils generate weak curvatures, which we partially compensate with a weakly focused laser beam in the vertical direction.  The first non-negligible terms of the resulting potential around the initial position of the atoms are $V_{res}(x,y,z)\simeq \frac{1}{2}m(-(2\pi\times 3.22 $Hz$)^2z^2+(2\pi\times 11 $Hz$)^2y^2-(12 $Hz$^2/\mu $m$)x^3)$. Here the axes are the same as in Fig.1. The anti-trapping curvature along $z$ is caused by non-perfect Helmoltz configuration of the Feshbach coils. In the $x$ direction an off-center dipole trap cancels a magnetic field gradient along the same direction. The resulting potential has a cubic spatial dependence, flat around the atoms position to allow for a free expansion in the disorder. In the $y$ direction the same optical trap contributes to a weak trapping potential.
By noticing that the typical energy scale for the system is of several tens on nK, we could define a spatial region in which the spurious fields stay below $\simeq$5nK, so that if the system remains within this region, we can consider negligible the effect of the residual curvatures. The size of the region amounts to 144$\mu$m along $z$, 42$\mu$m along $y$ and 112$\mu$m along $x$. In the fast diffusive regime (very small $V_R$), we indeed observe some deviations from the expected expansion when the cloud size approaches the region's boundaries. We expect that the effect of such a background potential is minimal at the mobility edge, where the typical energies we explored in this work range from 20nK to 100nK. Note also that the potential corresponds to antitrapping in two directions, and to trapping in the third direction, suggesting that the net effect on the 3D problem is less than that in the individual directions.

\section{Modulation spectroscopy: the coupling coefficient}
Using the numerical simulations, we try to estimate the dependence of the excitation probability $p$ on the initial energy of the atom $E$, at least for low modulation frequencies (so as to stay in the low-energy regime where the simulations are reliable). We find a typical linear scaling $p(E)=p_0+cE$ with $p_0\simeq0.5$ and $c\simeq0.2$nK$^{-1}$. If we consider this dependence in the calculations for $n(E,\hbar \omega)$, we find only a small shift of $E_c$ with respect to the one obtained for $p(E,\omega)=$const. Actually, even a 3 times larger $c$ than the one we get in the simulations would not change $E_c$ by more than 2nK. This very weak dependence on the actual form of $p$ is due to the fact that the features of the spectra that are relevant to determine $E_c$ simply depend on the behavior of $p$ in a small range of energies, between $E_0$ and the peak in $n_0(E)$. We can therefore conclude that the approximation $p(E,\omega)=$const provides reliable results for the mobility edge.

\end{document}